\documentclass[12pt,preprint]{aastex}

\slugcomment{To appear in Astrophysical Journal}

\shorttitle{Cygnus X-3's Little Friend}
\shortauthors{McCollough et al.}

\begin{document}


\title{Cygnus X-3's Little Friend}


\author{M. L. McCollough and R. K. Smith}
\affil{Smithsonian Astrophysical Observatory, 60 Garden Street,
    Cambridge, MA 02138, U.S.A.}
\email{mmccollough@head.cfa.harvard.edu}

\and

\author{L. A. Valencic}
\affil{Johns Hopkins University, 3400 N. Charles St., Baltimore, MD 21218}

\begin{abstract}
  Using the unique X-ray imaging capabilities of the {\it Chandra} observatory, a 
  2006 observation of Cygnus X-3 has provided insight into a singular feature 
  associated with this well-known microquasar.  This extended emission, located 
  $\rm \sim 16^{\prime \prime} ~$ from  Cygnus X-3, varies in flux and orbital phase (shifted by 0.56 in phase) with Cygnus X-3, 
  acting like a celestial X-ray ``mirror". 
  The feature's spectrum, flux and time variations allow us to determine the location, size, density, and
  mass of the scatterer.  We find that the scatterer is a Bok globule located along our line of sight,
  and discuss its relationship to Cygnus X-3.  This is the first time such a feature has been identified
  with the {\it Chandra} X-ray Observatory.
\end{abstract}


\section{Introduction}

At a distance of 9 kpc \citep{pbpt}, Cygnus X-3 is an unusual microquasar in which a 
Wolf-Rayet companion \citep{vk1} orbits a compact object with an orbital period of 
4.8 hours. It is a strong radio source routinely producing radio flares of 1 to 
$\rm \sim 20~Jy$ \citep{we2}.  It has been shown to produce radio jets 
and  the radio emission correlates with both the soft X-ray and hard X-ray 
emissions \citep{ma,jmj,as,mm}.
In 2000 {\it Chandra} observations found extended X-ray emission that is 
believed to be associated with Cygnus X-3 \citep{wh}.  In this paper we analyze a 2006 
{\it Chandra} observations of Cygnus X-3 and reexamine the previous Chandra observations 
discussed in \citet{wh}.  In particular we take a careful look at the timing and spectral properties of 
the extended feature and how they related to Cygnus X-3.

\section{Observations and Analysis}

Between 1999--2006 Cygnus X-3 was observed six times by {\it Chandra} using the ACIS-S/HETG with a 
Timed Event (TE) mode and covered a range of source activity (see Table 1).  These observations provided 
grating spectra of Cygnus X-3 and on-axis zero order images which allow for spatial and spectral 
analysis of Cygnus X-3 and its surroundings.

The primary observation used for this analysis was a 50 ksec quenched state observation 
(OBSID: 6601 hereafter referred to as QS) obtained 
during a period of high X-ray activity.  At the time of the observation the RXTE/ASM (2-12 keV) 
count rates were $\rm \sim 25-30 ~cts~s^{-1}$, Swift/BAT (15-50 keV) had an average count rate of 
$\rm \sim 0.0 ~cts~s^{-1}$, and the Ryle radio telescope (15 GHz) showed radio fluxes of 
$\rm \sim ~ 1~mJy$, all values typical of a Cygnus X-3 quenched state \citep{as,mm,we}.  This 
observation has the longest duration of any of the observations and the feature has the greatest 
number of photons of any of the observations.

Since this analysis involves a region near a bright point source we also need to determine 
the impacts of pileup on the data.  This can be determined by looking at the average number of counts per detection island 
(9 pixels region: see the \citet{pog}) per observing frame.  We do this by taking a series of annuli centered on Cygnus X-3 for the highest count rate 
observation (QS).  Each annulus has a radial thickness of 2 ACIS pixels ($\rm \sim 1\arcsec$) with the readout streak regions excluded.  
We then sum the counts in each annulus and divide by the number of observation frames\footnote{The number 
of observing frames can be determined either by extracting the keyword TIMEDEL from the event file header and 
dividing the exposure time by its value or using the acisf06601\_000N001\_stat1.fits file found in the {\it Chandra} secondary data products which 
contains a record of all of the frames in the observation.  This file must also have applied all the time corrections and filtering that 
were applied to the event file and correct CCD selection needs to be made.} 
and the of number detection islands (the area of the annulus in pixels divided by 9 pixels).  In Figure 1 is a plot of 
counts/frame as a function of radial distance from Cygnus X-3, with the solid line represents the entire 
observation and the dotted line the times of peak count rate.  From \citet{jd} we have taken the counts/frame values 
for which one would expected pileup of 1\% , 5\%, and 10\% and plotted them in Figure 1.  We see that at radial distance of greater than  
$\rm \sim 8\arcsec$ pileup should not be an issue and will not impact our analysis.

In our analysis of these observations, we used version 4.3 of the CIAO tools.  The {\it Chandra}
data retrieved from the archive were processed with ASCDS version 7.7.6 or higher. All event
files were filter by their good times intervals and barycenter corrected using the CIAO tool {\it axbary}.

\subsection{The Feature and its Characteristics}

For each observation, we have extracted a zeroth-order image between $\rm 1-8 ~keV$ binned at the 
nominal ACIS-S resolution ($\rm \sim 0.49\arcsec $).  Figure 2 and 5 shows images for 
the QS observation. In each individual image, there is a bright, unresolved core 
with  strong radial point-spread function (PSF) wings and a strong scattering halo 
\citep{pbpt}.  In each observation the feature reported by \citet{wh} is  present at the same 
location.  An analysis of the QS observation shows the feature 
($\rm RA:20^{h}32^{m}27.1^{s}, 
DEC:+40\degr 57\arcmin 33.8\arcsec$) lies at an angular distance of $\rm 15.6\arcsec$ 
from Cygnus X-3 at an angle of $\rm 68.5\degr$  from the orientation of the 
observed radio jets of Cygnus X-3.  The feature is extended and was fit using CIAO/SHERPA with 
a 2D-Gaussian with axes of $\rm 3.6\arcsec$ and $\rm 5.5\arcsec$ and a 
position angle of $\rm 78.7\degr$, measured counterclockwise with the top of
the image being $\rm 0\degr$.  If the feature and Cygnus X-3 are 
at the same distance, then their separation is $\rm 2.2~D_9$ light-years 
($\rm D_9$: distance in units of 9 kpc). This feature is present in all observations of Cygnus X-3 done
throughout the {\it Chandra} mission (1999 to present).

\subsection{Temporal Behavior}

The feature is located amidst a high background region.  This high background is due to the telescope
PSF and a dust scattering halo.  The majority of the photons in the background have energies above 
2 keV and as such arise from scattering off micro-roughness in the telescope optics; see the discussion in 
\citet{pbpt}.  In order to examine the temporal variability of the feature we need to subtract the background
flux from region of feature.  To do this we used segments of annuli centered on Cygnus X-3 for the feature 
and background regions to extract the light curve for the feature and the background (see Figure 2).   The
annuli were chosen to maximize the number of counts for the feature as well as have the feature and
background regions sample the same radial region of the PSF/dust halo while avoiding the readout streaks.  
All data extraction was done with the CIAO tool {\it dmextract}.

\subsubsection{Phase Folded Analysis}

The phase-folded (1-8 keV) light curves of Cygnus X-3, the background region, and background 
subtracted feature region are shown in Figure 3.  As might be expected the background which is mainly due to
the scattered PSF emission demonstrates the same 4.79 hr orbital variation with the slow rise and rapid observed drop in 
Cygnus X-3.  This is also reinforced in Figure 4 (top panel) which show no observed lag in the 
cross-correlation between Cygnus X-3 and the background.  However, the feature surprisingly shows the same 
orbital modulation {\it but with a phase shift of $\sim$ 0.6}. Phase-selected images reveal how this feature varies relative 
to Cygnus X-3 (see Figure 5), which can be seen more dramatically in the movie which accompanies 
this paper (see online version).  This is also confirmed in the cross-correlation between Cygnus X-3 and 
the feature seen in Figure 4 (lower panel) which show a clear lag that peaks around 9460 seconds which 
corresponds to a phase lag of $\sim$ 0.55.

{\it Fitting Phase Folded Light Curves:}  To address questions of a possible issues with background
subtraction (over-subtraction) we took the light curve from feature annulus and fit it using the following
model:

\begin{equation}
\rm C_{lf}[i] ~=~ C_{bkgd}[i] + a*C_{bkgd}[shift(i,j)] .
\end{equation}

Where $\rm C_{lf}[i]$ is the count rate for the feature's region in phase bin i, $\rm C_{bkgd}[i]$ is the
count rate from background region (scaled to the feature's region size) in phase bin i, and the last term  
$\rm a*C_{bkgd}[shift(i,j)]$ is the count rate of the background region scaled by $\rm a$ and shifted by 
$\rm j$ in phase.  The fit parameters are  $\rm a$ and $\rm j$.  We used the IDL routine MPFIT \citep{cm} to
fit the light curve.  We did multiple fits of the light curve using a number of different  
phase binnings and found good fits for all binnings with consistent fit values (see Table 2).  Our best 
fit a gave $\rm a~=~ 0.29 \pm 0.01$ and $\rm phase~shift(j) ~=~ 0.56 \pm 0.02$ (see Figure 6).

As a further test we replaced the second term in Equation 1 with a constant term which was used as a 
fit parameter.  We found no acceptable fits for any of the binnings (see Table 2).  
This result is consistent with the feature varying with the same period as Cygnus X-3 but shifted in phase.

{\it Phase Image:}  Finally, as another way to check to see if background subtraction could be
an issue for each {\it Chandra} detected event, a ``Cygnus X-3" phase value was determined
from its arrival time.  Photons falling into certain phase ranges were broken
in separate ``phase" images.  These images were assigned a certain color and
combined to form a color coded phase image.  The bands were:
{\tt (red) 0.3-0.63},  {\tt (green) 0.63-0.96}, and {\tt (blue) 0.96-0.3}.  The image is 
shown in Figure 7, note the blue color of the feature.  This indicates that bulk of the photons
are arriving in the 0.96-0.3 phase range.  If the feature was constant we would expect it to be white
(equal amounts of each color).  It is also important to note that
no background subtraction was done in this approach and hence there is no issue with the
background subtraction creating a false time/phase variation of the feature.

\subsubsection{Longer timescale Variability}

To examine the temporal behavior of the feature on longer timescales we determined the flux\footnote{The 
spectral fits and corresponding fluxes estimates are given and discussed in section 2.3 and 3.1.} of 
Cygnus X-3 and the feature for each of the {\it Chandra}/ACIS observations (see Table 1).  
When Cygnus X-3's flux is plotted versus the flux from the feature there appears to be a correlation (see
Figure 8) where  the dashed line is a linear fit to the data.  A Pearson's correlation test of the data 
yields a correlation coefficient of 0.98 indicating a linear correlation between the feature and Cygnus X-3.

\subsection{Spectrum}

The annuli used to extract the light curves (see Figure 2) were also used to extract spectra for the feature and
background region.  The final background-subtracted spectra of the feature contained from 78 to 3216 counts in 
the 1-8 keV band. Table 3 shows fits to the spectrum using simple absorbed power law and blackbody models all of 
which yield acceptable fits to the spectrum.   
 
All of the {\it Chandra}/ACIS observations were HETG grating observations.  The $\pm$ first order 
HEG spectra were combined and fit for each observation (see notes in Table 1).  The continuum for Cygnus X-3 is 
complex and dependent on the state of activity \citep{lh1,lh2,kk}.  In the X-ray (0.5-10 keV) the continuum 
can be modeled by partial covered disk blackbody during flaring/quenched states \citep{kk} and during the 
quiescence/transition states we found that continuum was best approximated by an absorbed power law.  In all cases 
a large number of spectral features were added (see Table 4) to improve the spectral fit.

We note that all of the spectra of the feature are more absorbed and very steep/soft, relative to the corresponding 
Cygnus X-3 spectra (see Table 3 \& 4).   As Cygnus X-3 transitions from a quiescent (hard state) to flaring/quenched 
state (soft state) its spectrum becomes softer.  The spectrum of the feature is shown to follow suit and becomes
 steeper/softer as Cygnus X-3's spectrum does.

\section{Nature of the Feature}

To understand the feature's nature the following must be explained:
{\tt (a)} it is clearly extended;
{\tt (b)} its flux varies in phase with Cygnus X-3 with a phase shift of 0.56;
{\tt (c)} the flux from the feature shows a correlation with the flux from Cygnus X-3;
{\tt (d)} the phase-averaged flux of the feature is  $\rm \sim 10^{-3}$ of Cygnus X-3's flux;
{\tt (e)} the time variation is 4.8 hrs but the separation between the feature
and Cygnus X-3 is at least $\rm 2.2 ~D_9$ light-years; and 
{\tt (f)} its spectrum is heavily absorbed and lacks hard X-ray flux relative to Cygnus X-3.

{\it Jet Emission:} It is natural to try to associate this feature with the Cygnus X-3 jet emission.  
However, Cygnus X-3 was in a quenched state throughout the {\it Chandra} observation and for several 
days before and after, during which jet activity is strongly suppressed.  Additionally, over a three 
year period prior to this observations Ryle/AMI-LA observed no radio flare exceeding 0.5 Jy \citep{gp}. 
Furthermore, in earlier {\it Chandra} observations where there is jet activity, the feature is 
fainter.  If the X-ray emission of the feature was due to the jet one would expect this to be 
synchrotron emission  and one would not generally expect the spectrum to be so steep and heavily 
absorbed.  Also the misalignment of the feature relative to the jets observed in the radio complicates 
this picture.

{\it Jet Impact Area:} Problems with this being a jet impact area have been noted by \citet{wh}
(location relative to the radio jets, jet precession, and jet collimation).  In
addition the observed flux correlation between the feature and Cygnus X-3
provides problems given a likely separation of $\rm 2.2~D_9$ light years.  
The strong phase modulation of the feature (varying the same as Cygnus X-3 by a factor of two), 
combined with a periodicity exactly matching that of Cygnus X-3, would be difficult to understand.  
One would expect the continuing impact of the jet would give rise to a brighter constant 
flux from the feature and drastically reduce the modulation which is observed.

{\it Wind Interaction:} The strongest arguments which were made by \citet{wh} for the feature's nature 
is that it is due to a wind/ISM interaction.  The feature's distance, flux 
correlation and phase modulation with Cygnus X-3 make such a model difficult to
reconcile with the observations.  In this interpretation the feature is created over timescales that
are long ($\rm \sim 2000~years$) compared to the observed orbital modulation.  Also the direct change 
in the flux of the feature with Cygnus X-3 flux is difficult to reconcile with such a model.  

\subsection{Scattering Solution}

A natural explanation for the feature's time variable behavior is that it 
is a result of scattering from a cloud (which acts as a kind of interstellar 
X-ray ``mirror") between Cygnus X-3 and the observer. This explanation would 
naturally lead to the observed phase difference between light curves as a 
difference in the path length of the scattered photons \citep{ts}.  This could also explain 
the flux correlation (see Figure 8) and overall flux difference ($\rm \sim 10^{-3}$) since 
only a small fraction of the total flux will be scattered toward the observer.  In figure 9 a
diagram of the expected geometry for the scattering from a cloud is shown.

The spectrum of the feature can also be modeled as a scattered version of Cygnus X-3's spectrum.  
At these energies scattering will modify the spectrum by $\rm A*E^{-2}$ \citep{rs}, due to the reduced scattering 
efficiency at higher energies \citep{jo}.  There will also be an additional reduction at low energies caused 
by absorption in the cloud ($\rm N_{cl}$) and multiple scattering.  
We can then model the feature's spectrum as due to scattering as

\begin{equation}
\rm S_{lf} ~=~  e^{-\sigma(E)N_H(lf)} *  A*E^{- \alpha} * S_{cont}.
\end{equation}

Where $\rm S_{lf}$ is the scattering model for the feature, $\rm e^{-\sigma(E)N_H(lf)}$ is the additional 
absorption due to the cloud (modeled using phabs in XSPEC),  $\rm A*E^{- \alpha}$ represents the high energy 
attenuation due to scattering (modeled using plabs from XSPEC), and $\rm S_{cont}$ is Cygnus X-3 continuum model
determined from the grating data (see Table 4).  The resulting fit parameters for the observations 
are shown in Table 5 and the fit to the QS observation's spectrum is shown in Figure 10. 
The resultant fluxes $\rm (1-8 ~keV)$ for the feature and Cygnus X-3 are given in Tables 4 \& 5.  The
flux ratio of the feature to Cygnus X-3 are all in the range of $\rm (4-6) \times 10^{-4}$.

This gives rise to a natural explanation of the observed spectral 
differences.  The loss of the flux below 2 keV is the result of additional 
absorption in the cloud and the reduced flux at higher energy simply reflects 
the drop in scattering efficiency.

{\it Distance to and Size of the Feature:} Assuming that this feature is due to an individual cloud 
the time delay is determined by the distance to the cloud \citep{ts,pk,pbpt}.   The resulting time 
delay can be written as: 

\begin{equation}
\rm \Delta t ~=~ \frac{\Theta^2}{2c}\frac{Dx}{1-x} ~=~ 1.15 \Theta^2 \frac{Dx}{1-x} .
\end{equation}

\noindent 
Where $\rm \Delta t$ is the time delay (in seconds), D is the distance (in kpc) to the source, 
$\rm \Theta$ is observed angular distance (in arcsec) from the source, x is the fractional distance 
of the scatter to the observer (see Figure 9), and c is the speed of light.  From the observed phase 
observed phase offset, we know that time delay is given by $\rm \Delta t = (0.56 + n) t_{cx3}$
where $\rm t_{cx3} = 17.25~ksec$ is the observed orbital period of Cygnus X-3, but we have an  
ambiguity in the total number orbital period offsets (n).  Using $\rm \Delta t ~=~0.56*t_{cx3}~=~9.66~ksec$ , 
D = 9 kpc, and $\rm \Theta = 15.6\arcsec$, the resulting fractional distance is x = 0.79.  This means that 
the cloud is close to Cygnus X-3 (within 1.9 kpc) and if $\rm n > 0$, this distance could be less.  

This degeneracy (n) can be removed using the fact that the feature is extended.  From equation (3) we would
expect that the delay scattered photons experience increases as a function of angle from
Cygnus X-3.  If we take the inner and outer edges of the feature to be $\rm 13.8\arcsec$ and 
$\rm 17.4\arcsec$ respectively then for x = 0.79 we would get a time delay of $\rm \sim 1.2 ~hrs$
across the feature.  For locations closer to Cygnus X-3 we would expect the delay across the feature to 
increase.  To test this we have taken the extraction and background annuli (see Figure 2) and divided into 
an inner and outer set of annuli ($\rm 11.2\arcsec$ to $\rm 15.6\arcsec$ and $\rm 15.6\arcsec$ to 
$\rm 20.0\arcsec$ respectively).  In Figure 11 we show a cross correlation of the inner and outer light curves 
(for the QS observation with 5 minute time resolution)
which shows a significant lag (at the 99 \% confidence level) at $\rm 0.9~hrs$.  This would correspond to the $\rm n = 0$ 
case. In Figure 12 are the phase folded light curves of the inner and outer region in which one can see that 
the light curve for the outer region is lagging the inner by $\rm \sim 0.2$ in phase.  This corresponds to 
a lag of $\rm \sim 1~hr$.  This puts the feature at a distance of 1.9 kpc from Cygnus X-3 making the observed 
dimensions of the feature 0.12 by 0.19 parsecs.

{\it Density of the Cloud:}   From the spectral fit to the feature for the QS observations we have an absorbing 
column density of the feature of $\rm 5.0^{+2.0}_{-1.7} \times 10^{22}~cm^{-2}$.  This value is consistent with column 
density determined from spectral fits to the other observations (see Table 5).  The column density of the feature 
can also be estimated from the flux ratio of the feature to Cygnus X-3 using the following relationship 
(see Appendix A for derivation)

\begin{eqnarray}
\rm \frac{F_{lf}}{F_{cx3}} &=& \rm N_H(lf) \left[\frac{\pi \tan \alpha_1 \tan \alpha_2  \cos^2 (\theta_s -\theta)}{(1-x)^2} \right] \nonumber \\
&& \rm \times \left[\sum\limits_{i=g,si} N^{i}_d  \int^{E_2}_{E_1}S(E)e^{-\sigma(E)N_H(lf)}  \!\!\!\int^{a_{max}}_{a_{min}} a^{-3.5} 
\left(\frac{d\sigma_s(E,a,\theta_s)}{d\Omega}\right) dadE  \right].
\end{eqnarray}

The flux ratio is equal to the product of three quantities.  The first ($\rm N_H(lf)$) is the column density of the feature, 
the second  is a solid angle term which relates to what fraction of Cygnus X-3's flux is intercepted by the feature, and final 
term is a scattering term which is a measure of the flux scattered by the dust in the cloud (this is solved for by numerical integration 
over the energy range and grain distribution).  The last two terms are depend on x, the fractional distance 
between the observer and Cygnus X-3.  Figure 13 shows a plot of the last two terms and their product as function of x.  The 
solid line is for scattering due to silicates and the dotted line for scattering due to graphite.  Equation 4 can be solved for 
the column density of the feature as a function of x as is shown in Figure 14. The parameters used to create these curves are given in Table 6.  
Also included are lines representing the column density from the spectral fit of QS with its uncertainties and a vertical line representing its 
location from the observed time delay.  What we find is good agreement with the column density determined from the spectral fits and the 
column density ($\rm N_{lf} = 3.6 \times 10^{22} ~cm^{-2}$) necessary to produce the observed flux ratio.  

If we assume that path-length along the line-of-sight through the cloud is similar to the observed dimensions of the feature 
[$\rm (3.7-5.9)\times 10^{17} ~cm$] we can make an estimate of the density of the feature.  Taking the column density to be 
$\rm 5.0^{+2.0}_{-1.7} \times 10^{22}~cm^{-2}$ we arrive at a density range of $\rm (0.6-1.9)\times 10^{5} ~cm^{-3}$ making the 
feature a dense molecular cloud.

{\it Mass of the Cloud:} From the estimate of the density of the cloud and the size determined from
the X-ray measurements we can make an estimate of the mass of the cloud.  Using the simplifying
assumption of a spherical cloud with a diameter of between $\rm 0.12-0.19~pc$ with a density of 
$\rm 10^5 ~cm^{-3}$ we arrive, from X-ray observations alone, at an estimate of the mass of the cloud to 
be $\rm 2 - 24 ~M_\sun$.  

From the cloud's size, density, and mass the feature has all of the characteristics of a Bok Globule
\citep{br, cyh},  but instead of seeing this as a dark obscuring feature in the optical we see it
shining in scattered X-rays.

\section{Relationship to Cygnus X-3}

What is the relationship of this feature to Cygnus X-3?  Three 
possibilities present themselves:  

(1) {\tt Random Alignment}: Cygnus X-3 and the
feature both lie in the Galactic plane 
($\rm l_{ii} =  79.845\degr,~  b_{ii} = 0.700\degr$).  The Cygnus X region that hosts 
Cygnus X-3 is rich in molecular clouds \citep{sn}.  So this may be just a chance alignment.  If so
this gives us insight into the nature and structure of molecular clouds in the ISM.  
In this case we would be looking across three Galactic arms (with Cygnus X near the Local Spur, the 
feature in the Perseus arm at $\rm \sim 5 ~kpc$, and Cygnus X-3 in the Outer Arm at $\rm 7-9 ~kpc$). 
Bringing Cygnus X-3 closer, to 7 kpc, does not greatly change the distance estimate to the feature 
or the need for three star forming regions along the line of sight. 

(2) {\tt Supergiant Bubble Shell}:  These structures have been observed in other
galaxies \citep{ks}.  They have typical radii of 0.5-1.0 kpc and are driven by the
radiation and outflows from OB associations, supernovae and their remnants, and XRBs. 
Molecular clouds have also been found to exist in these shells \citep{yr}.  Cygnus X-3 is a 
high-mass X-ray binary and likely still resides in such an OB 
association.  This gives a
natural explanation of the feature's location along our line-of-sight.  
Given the distance of the feature from Cygnus X-3 this would be a bubble comparable to the HI shell 
found in NGC 6822 \citep{dbw}.

(3) {\tt Microquasar Jet-Inflated Bubble}: Within NGC 7793 a powerful microquasar is driving 
a 300 pc jet-inflated bubble \citep{psm}.  Cygnus X-3 is a microquasar whose jets appear to be  
aimed along our line of sight \citep{ma,jmj}.  It is possible that rapid cooling near the working 
surface of the jet, in the shell of the cocoon, may allow a dense molecular cloud to form.  
This would explain the nature of the feature as well as its alignment with Cygnus X-3.  
Although it should be noted that there is research which suggest that the jet may not be close to the 
line of sight \citep{mpp}, which would make this a less likely option.

Finally it should be noted that a combination of both (2) and (3) may be possible.  The radiation and outflows 
from the OB association may create a large low density cavity in which the microquasar jet can more easily 
propagate.  Evidence for large-scale cavities surrounding other microquasars has been noted \citep{hz}.  This could explain the large 
distance of the feature from Cygnus X-3 (1.9 kpc) and reduce the 
energetics necessary to produce it.  The feature may be located at the place where jet interacts which the 
wall of the cavity.

\section{Conclusion}

This feature and its temporal relationship to Cygnus X-3 have unveiled the unique 
interaction between a microquasar and its environment.  It has given us a tool to 
probe the nature and structure of molecular clouds, providing information on their size and 
shape, possibly due to the microquasar interaction or the 
presence of ordered magnetic fields in the ISM.  It has also given us our first X-ray view of a Bok
Globule.

To date this is the first such feature found with {\it Chandra}.  If the feature is indeed due to a
microquasar interacting with its environment then we would expect there to be very few.  
This would be due to the limits of small angle scattering in the X-ray and the need for the lobes and 
associated molecular clouds to be aligned close to our line of sight.  Depending on the nature of these sources the best 
candidates would be high mass X-ray binaries (because of their young age and hence likely relationship with an OB 
association and star forming regions) with a relatively short orbital periods.


\section{Acknowledgments}

MLM wishes to acknowledge support from NASA under grant/contract G06-7031X and NAS8-03060. MLM would
also like to acknowledge the useful discussions with Ramesh Narayan concerning the scattering path
through and interactions with interstellar clouds.  We wish to thank the referee for the helpful comments and 
suggestions.  This research has made use of data obtained from the Chandra Data Archive and software provided 
by the Chandra X-ray Center (CXC).

\appendix

\section{Derivation of Scattering Relations for Cygnus X-3's and the Feature}

It is possible to derive some of the scatter's properties by comparison of the fluxes of Cygnus X-3 and the feature.  
This derivation is similar to that done for the scattering halo intensity of \citet{sd}.  The geometry being used can be found in Figure 9.
If we take the unabsorbed luminosity of Cygnus X-3 as a function of energy to be $\rm L_{cx3} (E)$ then the X-ray 
luminosity at the feature is given by 

\begin{equation}
\rm L_{lf} (E) = \frac{L_{cx3}(E) e^{-\sigma(E)N_H (r_s)}}{{4 \pi r_s}^2} ~.
\end{equation}

Where $\rm \sigma(E)$ is the X-ray absorption cross section, $\rm N_H (r_s)$ is column density along the path 
($\rm r_s$) between Cygnus X-3 and the cloud.

The photons scattered into a solid angle $\rm d\Omega$ by a single
scatter for a source of luminosity $\rm L_{s}(E)$ is given by

\begin{equation}
\rm P (E) = L_{s} (E) \frac{d\sigma_s}{d\Omega} d\Omega ~.
\end{equation}

Where $\rm {d\sigma_s}/{d\Omega}$ is the differential scattering cross section (see \citet{ml,mg}).

For a telescope with collecting area $\rm A^{'}$ we can chose  
$\rm d\Omega$ such that $\rm A^{'} = {r_o}^2 d\Omega$,
where $\rm r_o$ is the distance from the scatter to the observer. 

The photon count rate that the observer will detect from scattering from a single
dust particle as

\begin{equation}
\rm C_s (E) = L_s (E) e^{-\sigma(E)N_H (r_o)} \left( \frac{d\sigma_s}{d\Omega} \right) \frac{A^{'}}{{r_o}^2} 
= \frac{L_{cx3}(E) e^{-\sigma(E)[N_H (r_s) + N_H (r_o)]}}{{4 \pi r_s}^2} \left( \frac{d\sigma_s}{d\Omega} \right) \frac{A^{'}}{{r_o}^2} ~.
\end{equation}

Where $\rm N_H (r_o)$ is the column density between the scatter and the observer.

The feature has been fit with an elliptical Gaussian with semi-major and semi-minor axes of $\rm r_1$ and $\rm r_2$ respectively.  
For a distance of $\rm xD$ one can use the angular measurements of the semi-major and semi-minor axes, $\rm \alpha_1$ and $\rm \alpha_2$ 
respectively, to give the surface area of the feature as $\rm A_{lf} = \pi r_1 r_2 =\pi x^2 D^2 \tan \alpha_1 \tan \alpha_2$.
For a scattering region thickness of $\rm l_s$ the scattering volume of the feature is given by $\rm V_{lf} = A_{lf} l_s$.
For a dust grain number density of $\rm n_g$ we can determine the total count rate and flux for the feature as

\begin{equation}
\rm F_{lf}(a,E) = \frac{C_{tot}(E)}{A^{'}} = 
\frac{n_g(a) L_{cx3}(E) e^{-\sigma(E)[N_H (r_s) + N_H (r_o)]}}{{4 \pi r_s}^2} \left(\frac{d\sigma_s}{d\Omega}\right) \frac{A_{lf} l_s}{{r_o}^2} ~. 
\end{equation}

In general $n_g$ will be a function of dust radius $\rm a$.  For our case we will consider $\rm n_g$ to be uniform spatially at the location of the feature. 
Integrating over the dust size distribution gives the flux for the feature as

\begin{equation}
\rm F_{lf}(E) = \int^{a_{max}}_{a_{min}} \frac{n_g(a) L_{cx3}(E) e^{-\sigma(E)[N_H (r_s) + N_H (r_o)]}}{{4 \pi r_s}^2} 
\left(\frac{d\sigma_s}{d\Omega}\right) \frac{A_{lf} l_s}{{r_o}^2}da ~.
\end{equation}

We can simplify the above equation by noting that the observed angle $\rm \theta$ for the feature is very small and the overall angular dimensions 
( $\rm \alpha_1$ and $\rm \alpha_2$) of the feature are small.  In this case the path traveled by the scattered photon
will be very close to the path traveled by the unscattered photon. Because of this the total hydrogen column density for both paths should be the same except for the additional 
column density of $\rm N_H(lf)$ along the scattered path due to the feature.  If we take $\rm N_H(cx3)$ to be the column density between the observer and Cygnus X-3 then the column
density along the path of the scattered photon can be written as

\begin{equation}
\rm N_H (r_s) + N_H (r_o) =  N_H(cx3) + N_H(lf)~.
\end{equation}

$\rm F_{cx3}(E)$ the observed absorbed flux from Cygnus X-3 can be written as

\begin{equation}
\rm F_{cx3}(E) = \frac{L_{cx3}(E) e^{-\sigma(E)N_H (cx3)}}{{4 \pi D}^2} ~.
\end{equation}

Finally from the scattering geometry (see Figure 9) we have 

\begin{equation}
\rm \frac{1}{r_s} =  \frac{\cos (\theta_s -\theta)}{D(1-x)} ~.
\end{equation}

Where $\rm x$ is the projected distance along path between the observer and Cygnus X-3.  The measure angle of the feature ($\rm \theta$) and the scattering angle ($\rm \theta_s$) 
are related to the projected distance $\rm x$ by $\rm \theta =  (1-x)\theta_s$.  Using the above relations we arrive a the following substitution\footnote{We note that this equation 
is similar to Equation A5 in \citet{sd}.  However, in that paper, one factor of $\cos(\theta_s - \theta)$\ was omitted in error, although this makes no difference to either their or 
our final results.}

\begin{equation}
\rm \frac{L_{cx3}(E) e^{-\sigma(E)[N_H (r_s) + N_H (r_o)]}}{{4 \pi r_s}^2} \approx \frac{F_{cx3}(E)e^{-\sigma(E)N_H(lf)}\cos^2 (\theta_s -\theta)}{(1-x)^2} ~.
\end{equation}

If we integrate over energy bandpass and  replace $\rm F_{cx3}(E)$ by $\rm F_{cx3}S(E)$ where $\rm F_{cx3}$ represents the measured flux from Cygnus X-3 and $\rm S(E)$ 
is its spectral form (normalized to one) we can write the flux relationship of the feature to Cygnus X-3 as

\begin{eqnarray}
\rm \frac{F_{lf}}{F_{cx3}} & = & \frac{\pi \tan \alpha_1 \tan \alpha_2 l_s \cos^2 (\theta_s -\theta)}{(1-x)^2} \nonumber \\
&& \rm \times \int^{E_2}_{E_1}S(E)e^{-\sigma(E)N_H(lf)}\!\!\!\int^{a_{max}}_{a_{min}} n_g(a)\left(\frac{d\sigma_s(E,a,\theta_s)}{d\Omega}\right) dadE ~.
\end{eqnarray}

Assuming a MRN grain size distribution \citep{mrn, wd} then we have

\begin{equation}
\rm n_g (a) =  n_h \sum\limits_{i=g,si} N^{i}_d a^{-3.5} ~.
\end{equation}

Where $\rm n_h$ is the hydrogen number density of the cloud, $\rm a$ is the radius of the grain and $\rm N^{i}_d$ is are the normalization in $\rm (grains/H~atom)/\mu m$ for graphite
(g) and silicates (si).  If we also
note that $\rm n_h l_s$ is simple the column density of the cloud $\rm N_H(lf)$.  Substituting $\rm n_g (a)$ in A10 gives us

\begin{eqnarray}
\rm \frac{F_{lf}}{F_{cx3}} & = & \rm N_H(lf) \left[\frac{\pi \tan \alpha_1 \tan \alpha_2  \cos^2 (\theta_s -\theta)}{(1-x)^2} \right] \nonumber \\
&& \rm \times \left[ \sum\limits_{i=g,si} N^{i}_d \int^{E_2}_{E_1}S(E)e^{-\sigma(E)N_H(lf)}  \!\!\!\int^{a_{max}}_{a_{min}} a^{-3.5} 
\left(\frac{d\sigma_s(E,a,\theta_s)}{d\Omega}\right) dadE  \right] ~.
\end{eqnarray}

\begin{figure}
\includegraphics[angle=90,scale=.7]{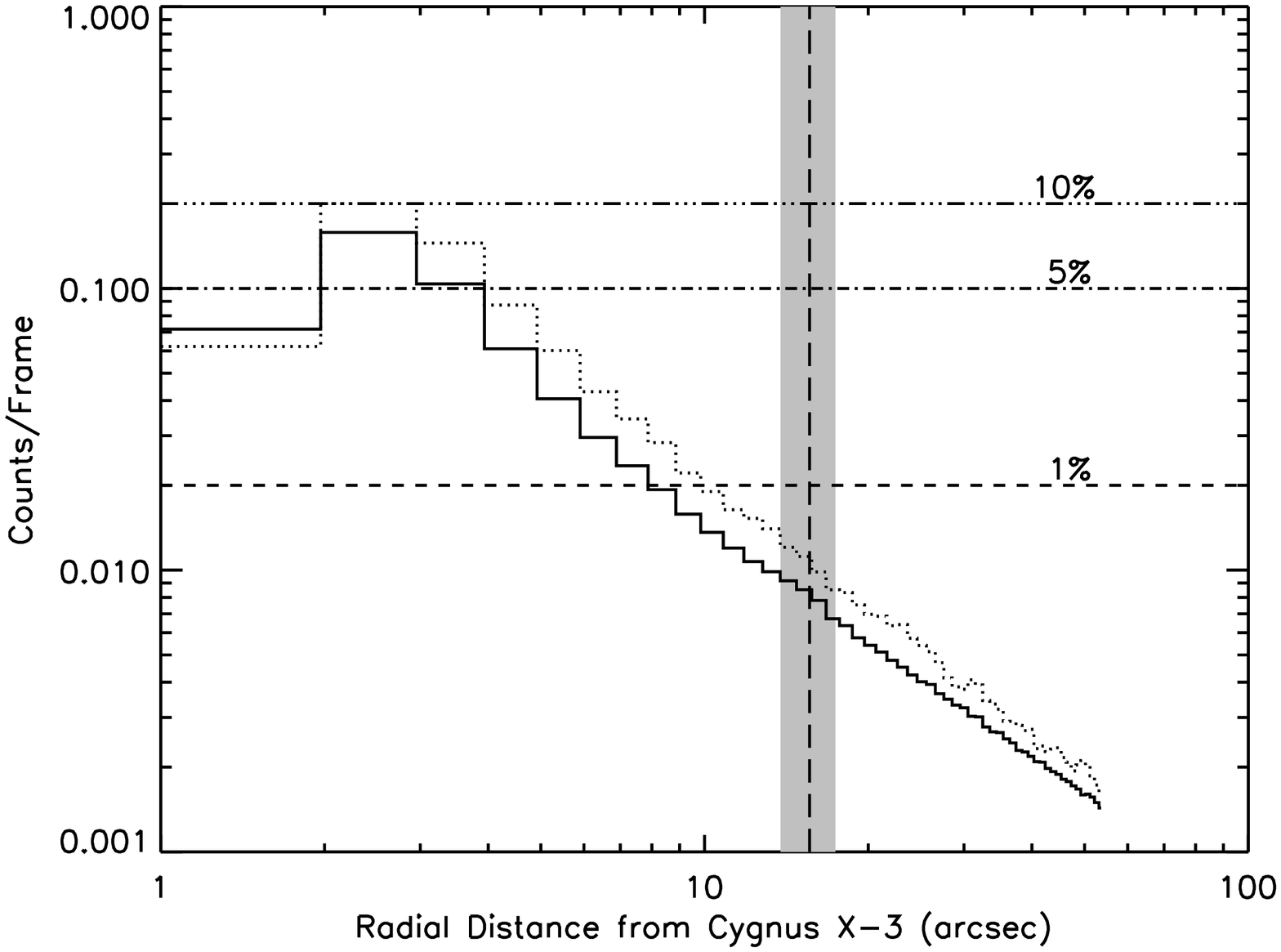}
\caption{Plotted are the counts/frame for the QS observation as a function of radial distance from Cygnus X-3.  The solid line is for the 
entire observation and the dotted line is for the times of the highest count rates (Cygnus X-3 phases 0.6-0.7).  The counts/frame for various levels of expected pileup are 
given \citep{jd}.  The long dashed vertical lines is the location of the feature and the shaded region give the feature's radial extent.  Pileup should not be an issue for 
the observations used in this analysis.}
\end{figure}

\begin{figure}
\includegraphics[angle=0,scale=0.75]{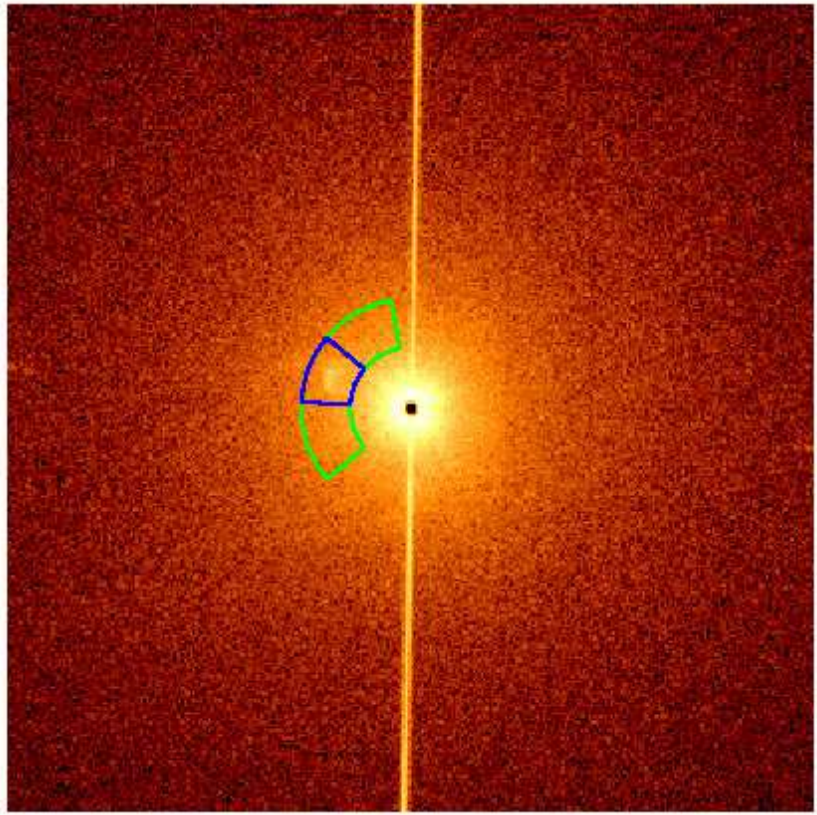}
\caption{ Segments of annuli centered on Cygnus X-3 were used for the feature and background regions 
to extract the light curve and spectrum  for the feature (blue) and the background (green).  The inner edge
of the annuli is $\rm 11.2\arcsec$ from Cygnus X-3 and the outer edge is $\rm 20.0\arcsec$.}
\end{figure}

\clearpage

\begin{figure}
\includegraphics[angle=90,scale=.70]{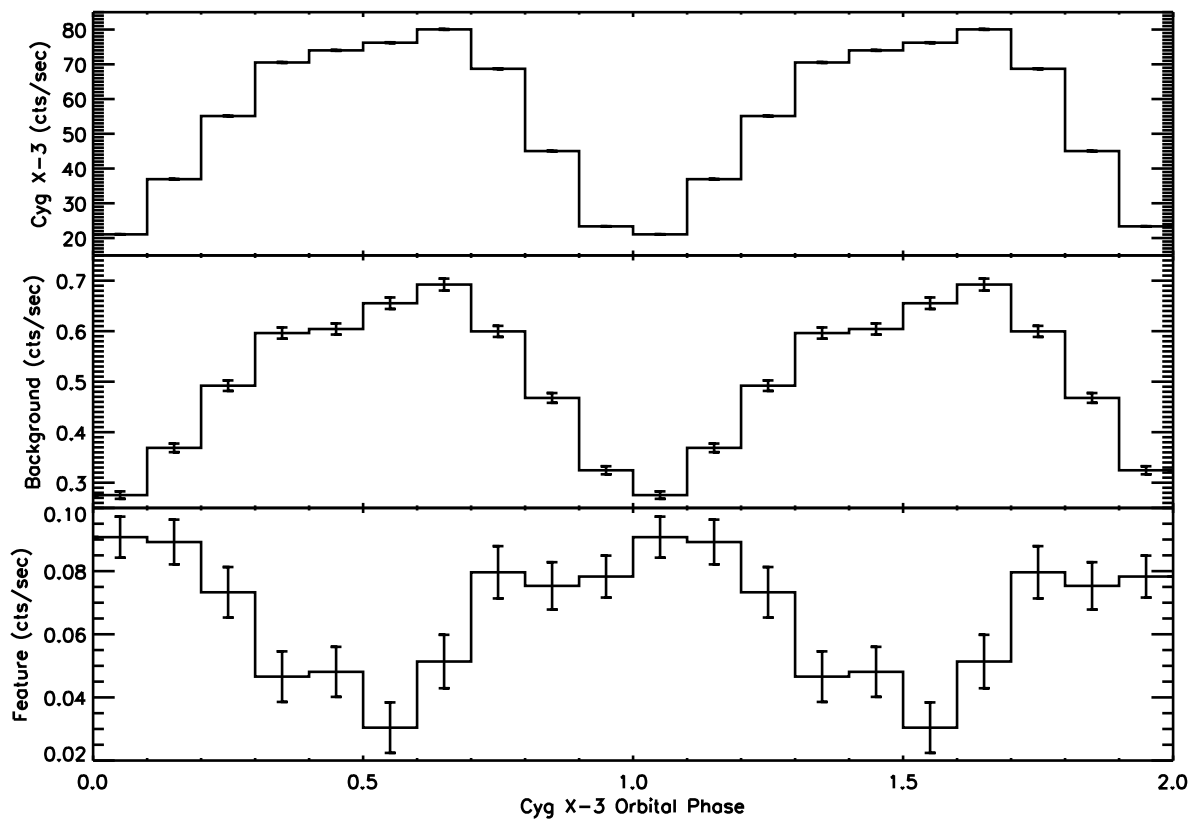}
\caption{ Phase folded 1-8 keV light curves of Cygnus X-3, the background, and the feature from the QS observation.  
{\it Top:} Phased fold light curve of Cygnus X-3.  The data were taken from the $\pm$ first order HEG and MEG spectra.
{\it Middle:} Phase folded light curve of the background region. 
{\it Bottom:} Phase folded light curve of the feature from QS (background subtracted). It can clearly be seen 
that the feature exhibits the same slow rise and rapid drop that one sees in Cygnus X-3 but with a phase 
lag of $\rm \sim 0.6.$}
\end{figure}


\begin{figure}
\includegraphics[angle=90,scale=0.70]{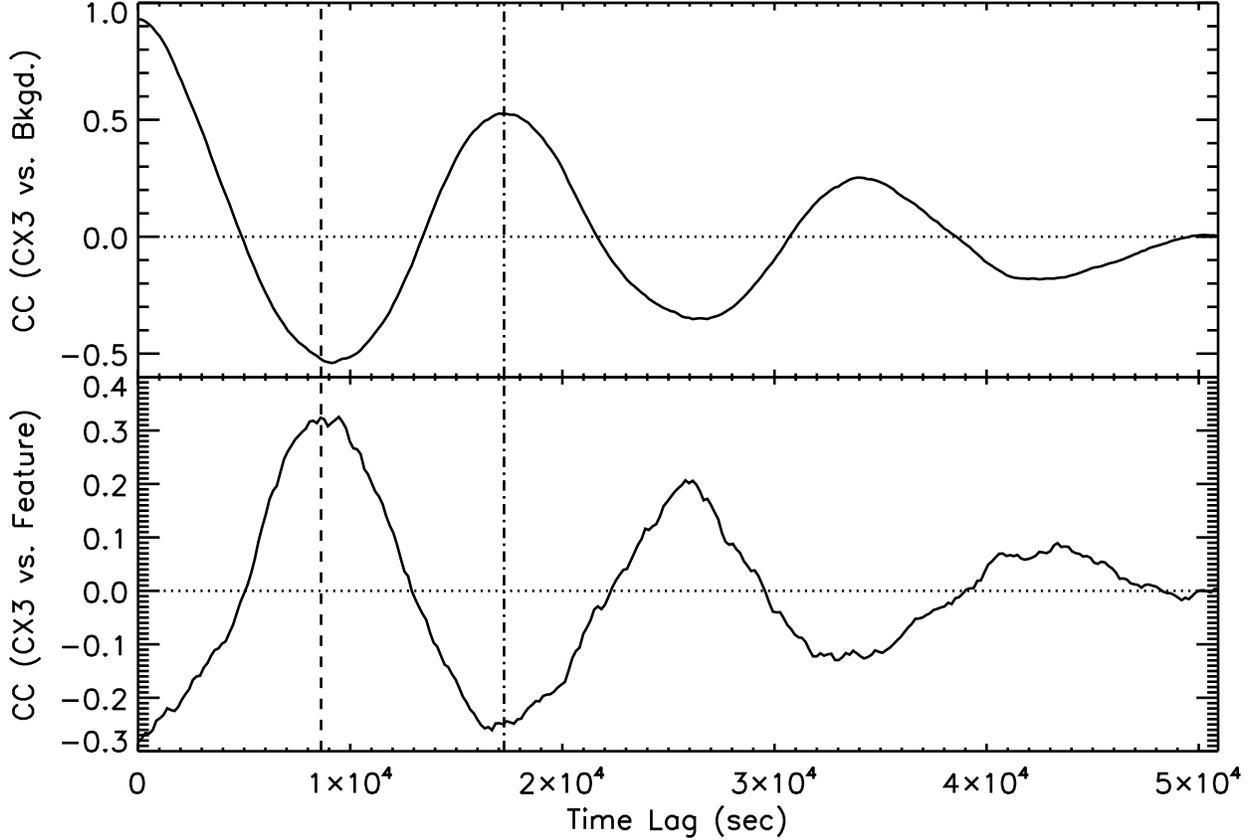}
\caption{ {\bf Cross correlation plots:}  The dashed vertical line corresponds to a lag of half of Cygnus X-3's
period and the dot-dashed corresponds to a lag of Cygnus X-3's period.  {\it Top:} The cross correlation 
between Cygnus X-3 ( grating data) and the background region (1-8 keV) using  600 second time samples.  Note 
there is no lag between them. {\it Bottom:} The cross correlation between Cygnus X-3 ( grating data) 
and the feature (background subtracted) using 172 second time samples ($\sim$ 0.01 of Cygnus X-3's
period). Note the lag of 9460 seconds which corresponds to a phase shift of $\sim$ 0.55.}
\end{figure}

\begin{figure}
\plottwo{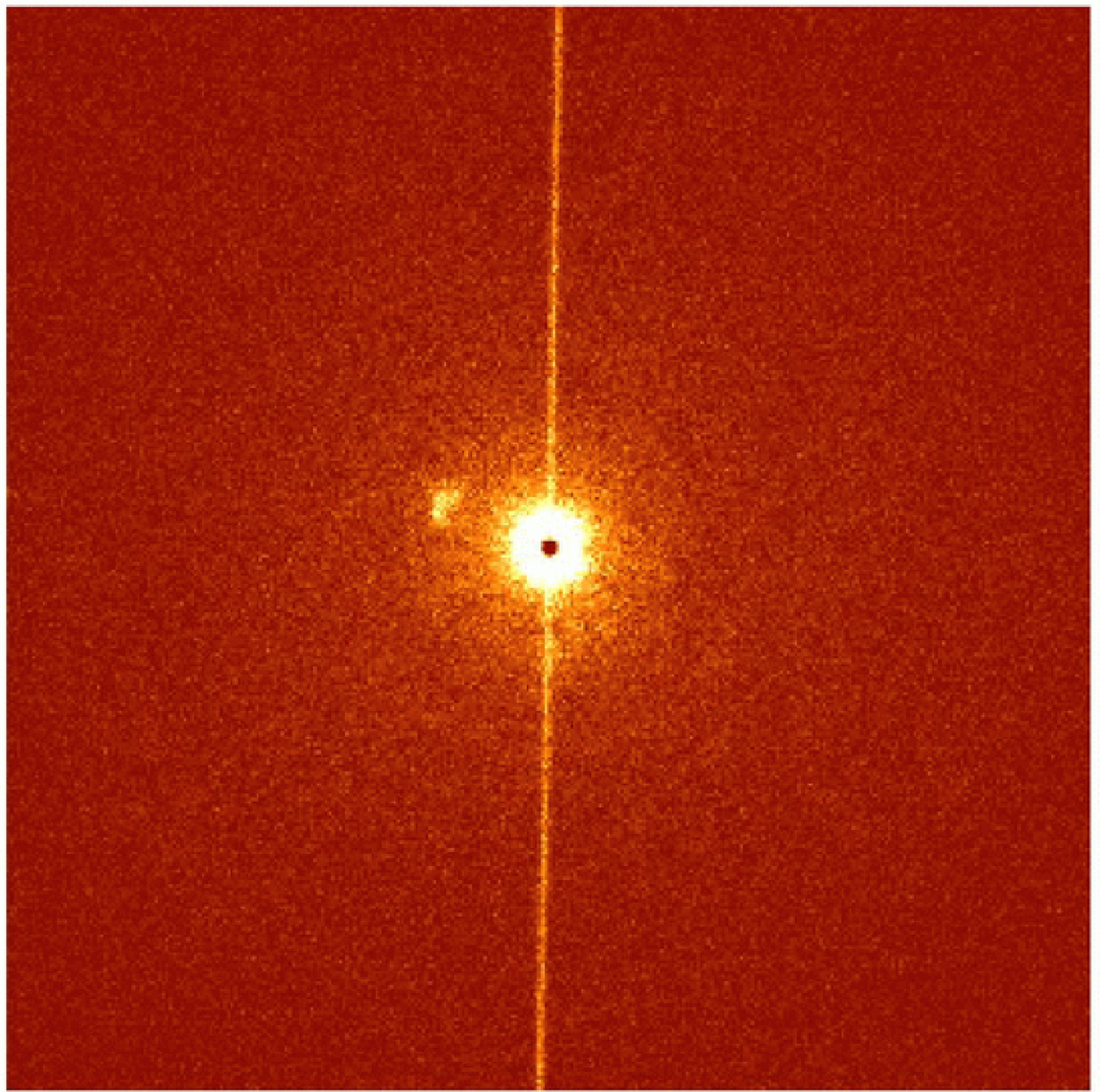}{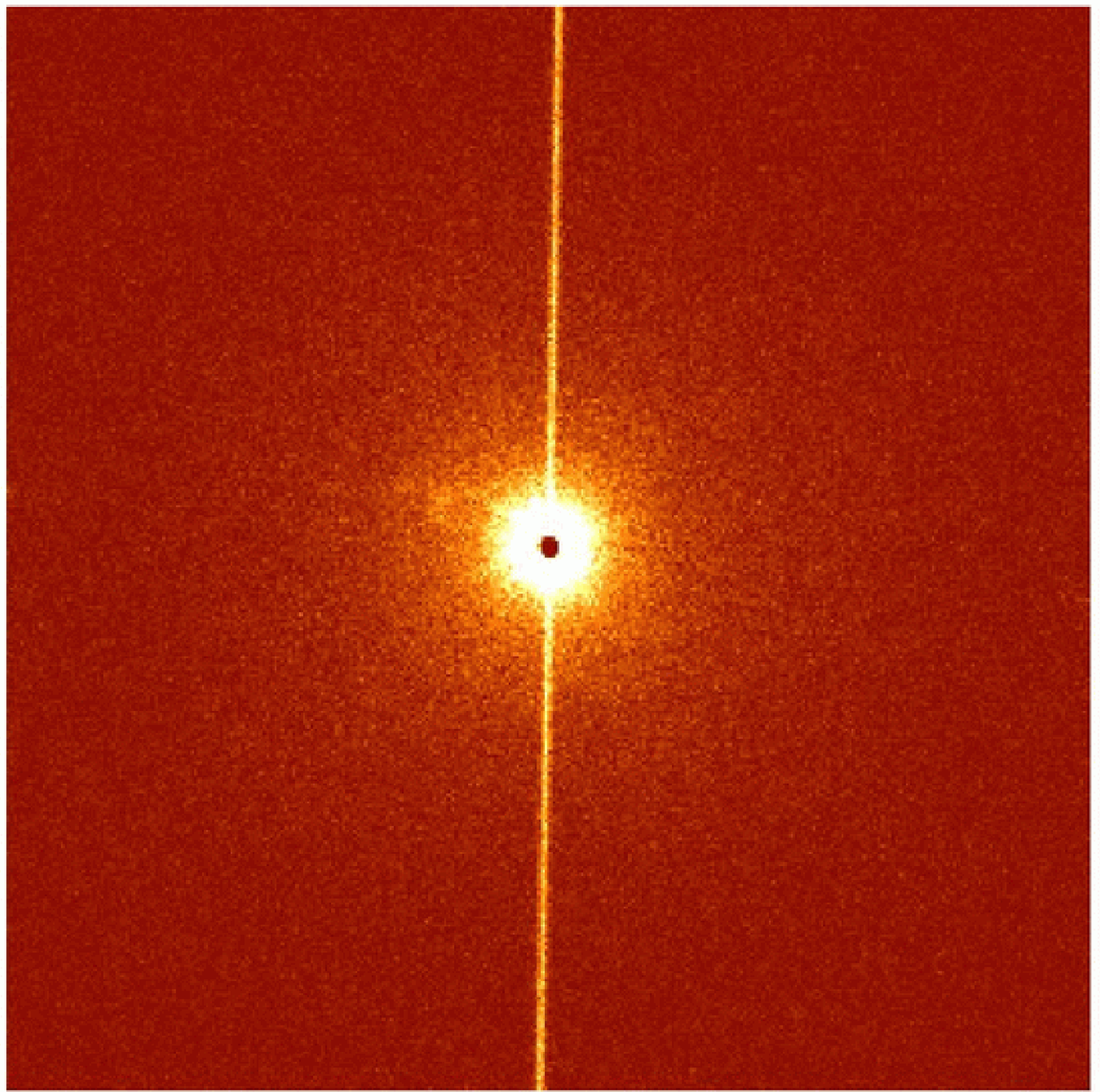}
\caption{These phase selected images were created from the zero-order image 
{\it Chandra} grating observation (QS) using photons with energies between 1-8 kev.  The count range and color scales 
are identical for both images.
The bright readout streak caused by the ACIS CCD readout and the ``cratering" of the central 
source due to pileup are both instrumental effects.
{\it left:} Phase range 0.96-0.3 image.  The feature is prominent.  
{\it right:} Phase range 0.5-0.8 image.  The feature is very weak.
{\it Online electronic edition:} To better visualize the phase relationship between the feature and Cygnus X-3 a movie 
was created.  For all of the events detected in QS a Cygnus X-3 orbital phase based on their arrival 
time was determined. The data was filtered to only include 1-8 keV energy photons.  For the image 
a 256 by 256 pixels region centered on Cygnus X-3 was used.  A set of images were created based on 
phase intervals, starting at 0.0 phase and with a duration of 0.20 of Cygnus X-3's phase.  Each image 
was successively shifted by a 0.01 of Cygnus X-3's phase until images for the full Cygnus X-3 period 
were created and compiled to form the movie.  The duration of the observation (50 ksec) corresponds to 
2.96 orbits of Cygnus X-3, so phase coverage is relatively uniform.  The duration of each image 
(0.2 phase) corresponds to an integration time of $\rm \sim 10^4~seconds$.}
\end{figure}

\clearpage

\begin{figure}
\includegraphics[angle=90,scale=.70]{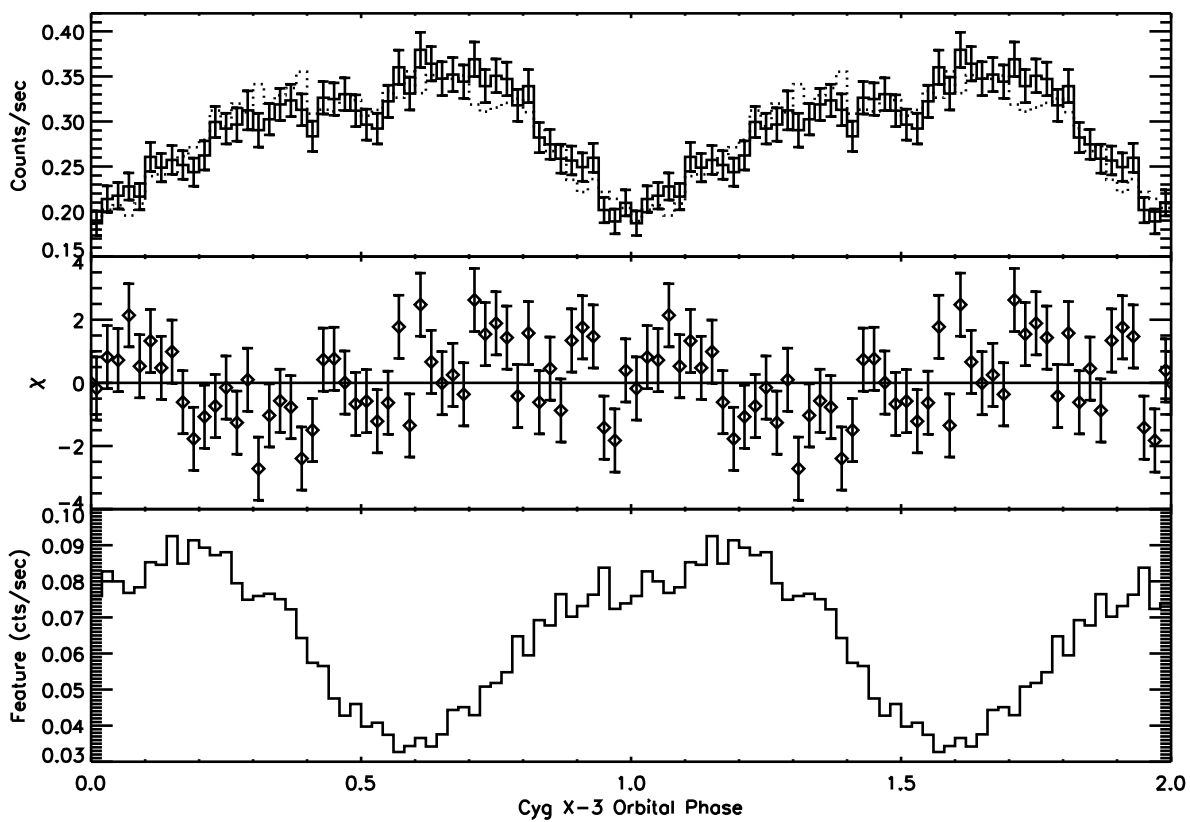}
\caption{Fits to the phase folded 1-8 keV light curves of the feature with no background subtraction.  
{\it Top:} Phase folded light curve of Cygnus X-3 (50 phase bins) with a fit to the light curve (dotted line).
{\it Middle:} Data minus model fit divided by the data error bars.  With $\pm ~1 ~\sigma$ error bars.
{\it Bottom:} The model of the phase folded light curve of the feature from the fit.}
\end{figure}


\begin{figure}
\includegraphics[angle=0,scale=.45]{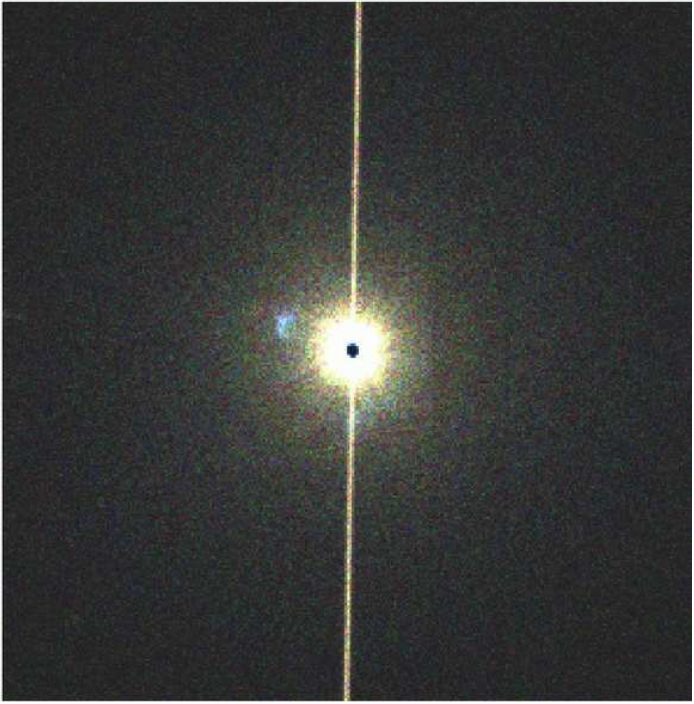}
\caption{A color coded phase image of Cygnus X-3 region.  
Note the blue color of the feature.  This indicates that bulk of the photons
are arriving in the 0.96-0.3 phase range.  It is also important to note that
no background subtraction was done and hence there is no issue with the
background subtraction creating a false time/phase variation of the feature.
}
\end{figure}

\begin{figure}
\includegraphics[angle=90,scale=0.75]{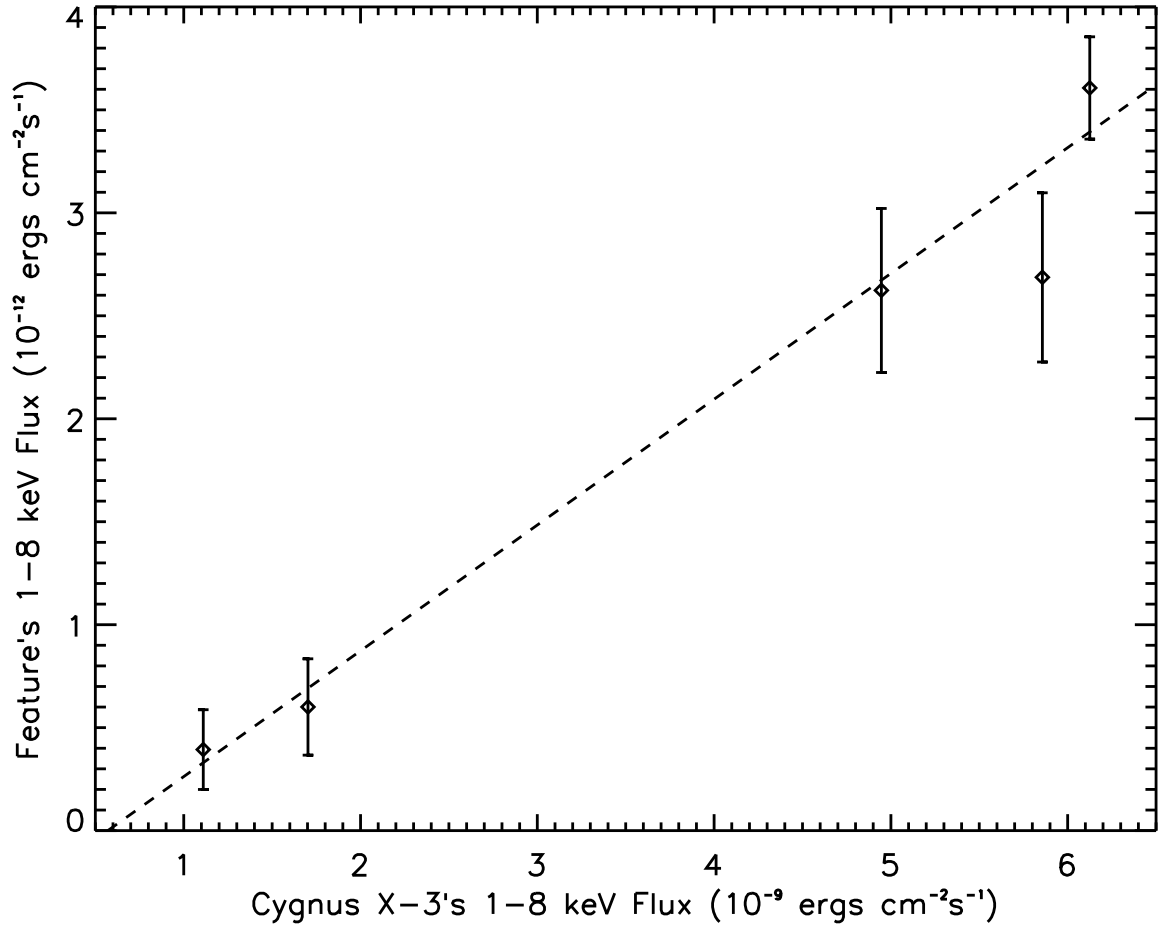}
\caption{Plotted is the $\rm 1-8~keV$ flux of Cygnus X-3 (determined from the grating data) versus 
the $\rm 1-8~keV$ flux of the feature for each of the ACIS observations using the scattering model.  The dashed line is a 
linear fit to the data.}
\end{figure}


\clearpage

\begin{figure}
\includegraphics[angle=0,scale=0.60]{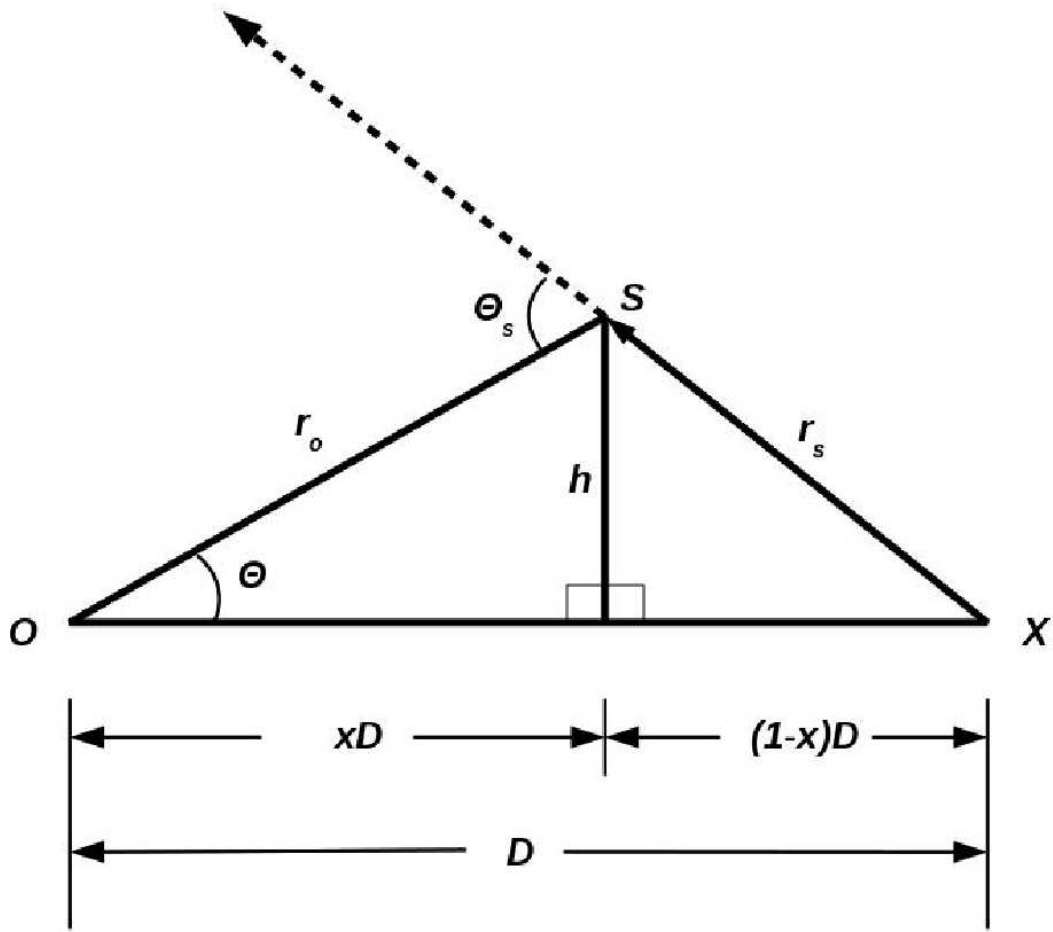}
\caption{ {\bf Scattering Diagram Definitions:}  Above is shown a diagram of how the scattering is 
taking place with various parameters labeled.
X~: Location of the X-ray source (Cygnus X-3).
S~: Location of the feature.
O~: Location of the observer.
D~: Distance from the observer to Cygnus X-3.
x~: Distance to the scatter along the path to Cygnus X-3.
$\rm \theta$~: Observed angle of the feature from Cygnus X-3.
$\rm \theta_s$~: The scattering angle of the X-rays from Cygnus X-3 by the feature.
$\rm r_s$~: Distance from Cygnus X-3 to the feature.
$\rm r_o$~: Distance from the feature to the observer.
h~: Distance from the the feature to Cygnus X-3's line of sight to the observer.}
\end{figure}

\clearpage

\begin{figure}
\includegraphics[angle=-90,scale=0.60]{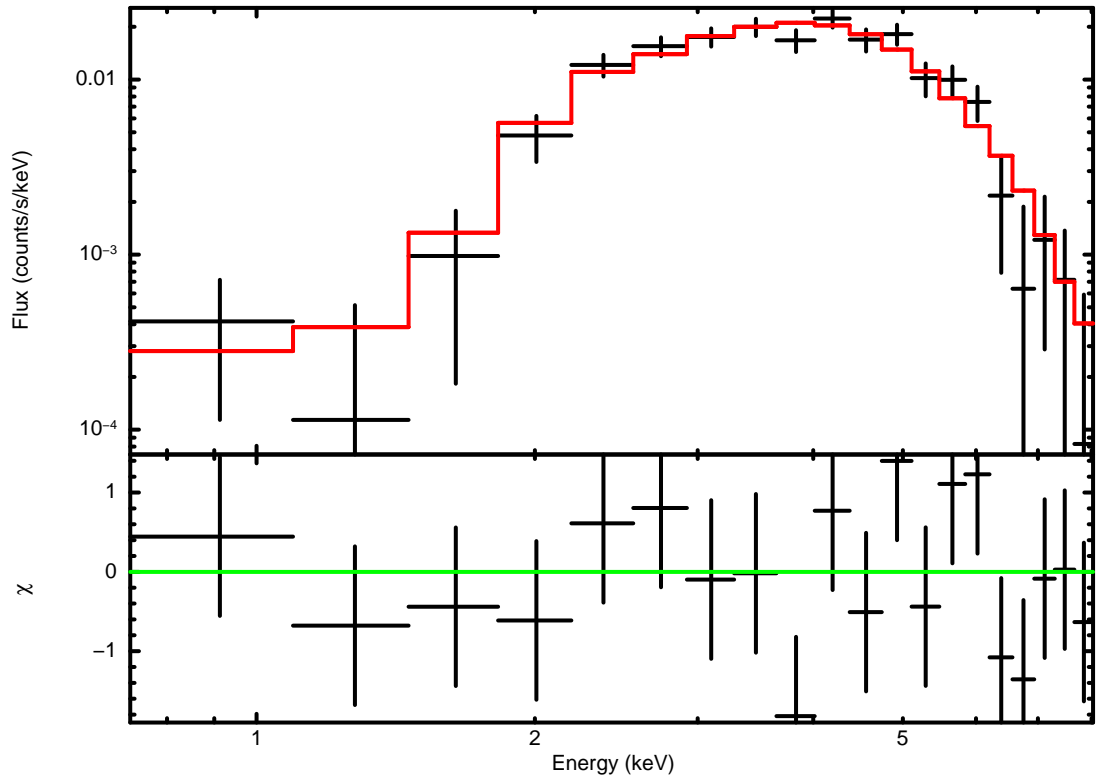}
\caption{ The extracted spectrum of the feature, taken from QS, is shown above.  Also given is the 
fit (in red) and residuals of a scattering model.  A good agreement is found between the data 
and model.}
\end{figure}

\begin{figure}
\includegraphics[angle=90,scale=0.75]{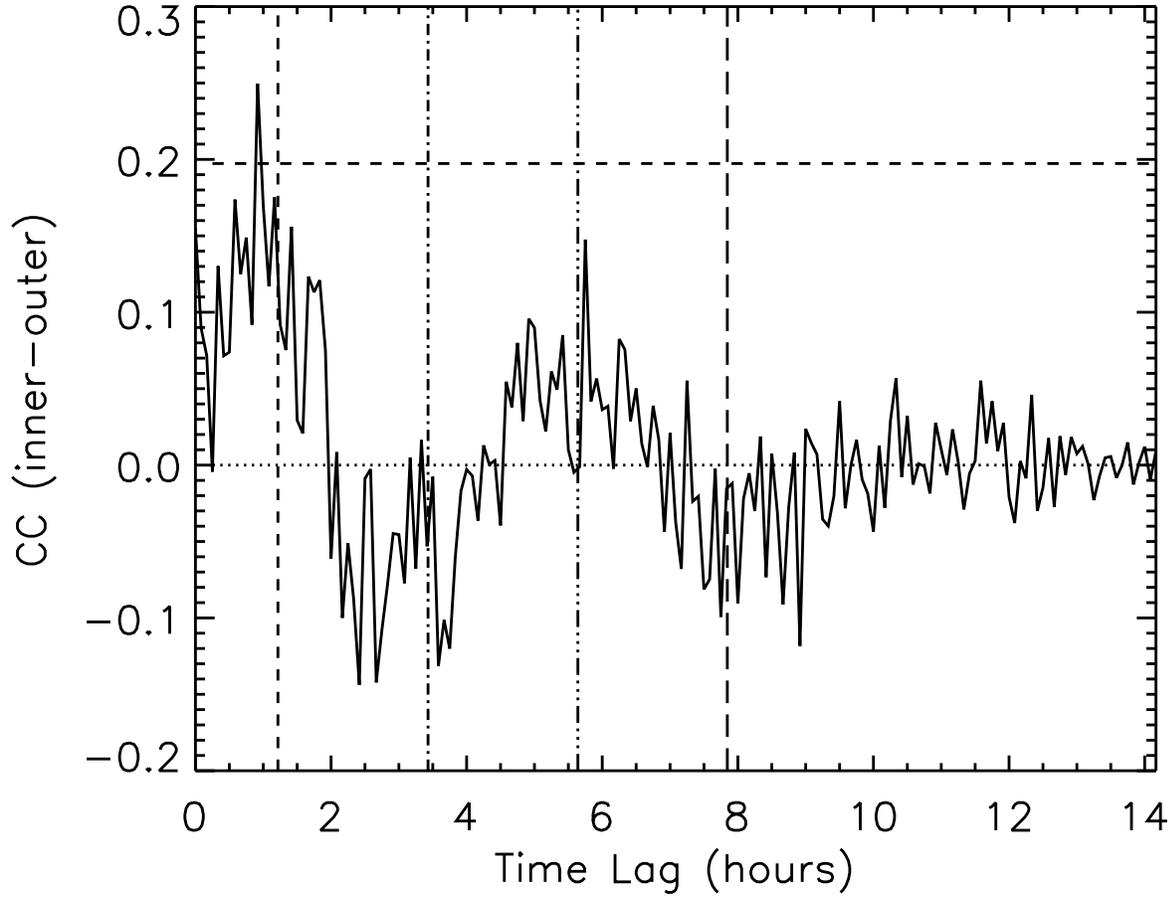}
\caption{A plot of the cross correlation between the inner 
($\rm 11.2\arcsec$ to $\rm 15.6\arcsec$) and outer ($\rm 15.6\arcsec$ to $\rm 20.0\arcsec$) regions
of the feature (both background subtracted).  The time resolution was 5 minutes for the light curves. The dashed horizontal lines corresponds 
to a 99\% confidence level.  The dash, dot-dash, 
dots-dash, and large dash vertical lines correspond to lags one would observe across the feature for 
n = 0, 1, 2, and 3 respectively.  Note the prominent peak at  $\rm \sim 1~hour$ corresponding to $\rm n~=~0$ lag.}
\end{figure}

\begin{figure}
\includegraphics[angle=90,scale=.7]{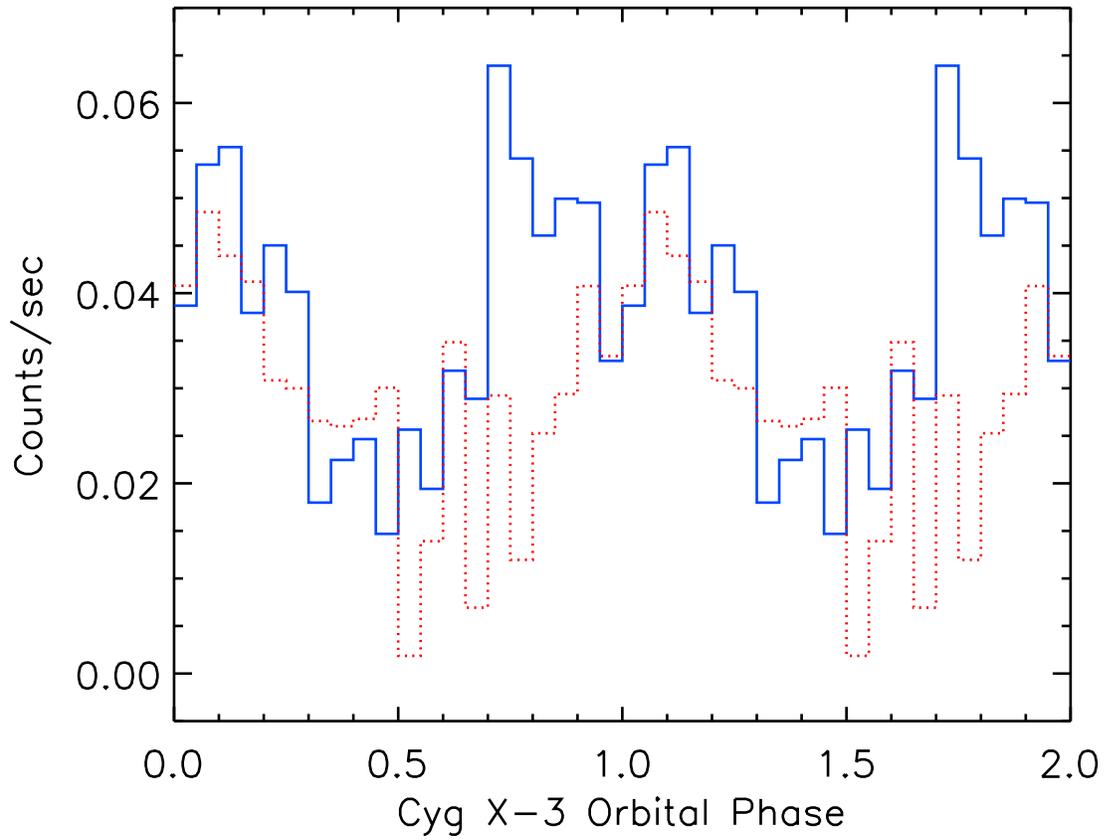}
\caption{A plot of the phase folded light curves of the inner (solid: blue online) and outer (dotted: red online) regions of the feature (both background subtracted). 
There is a noticeable lag in outer relative to the inner of $\rm \sim 0.2$ in phase which corresponds to $\rm \sim 1~hr$ in time.}
\end{figure}

\begin{figure}
\includegraphics[angle=90,scale=.7]{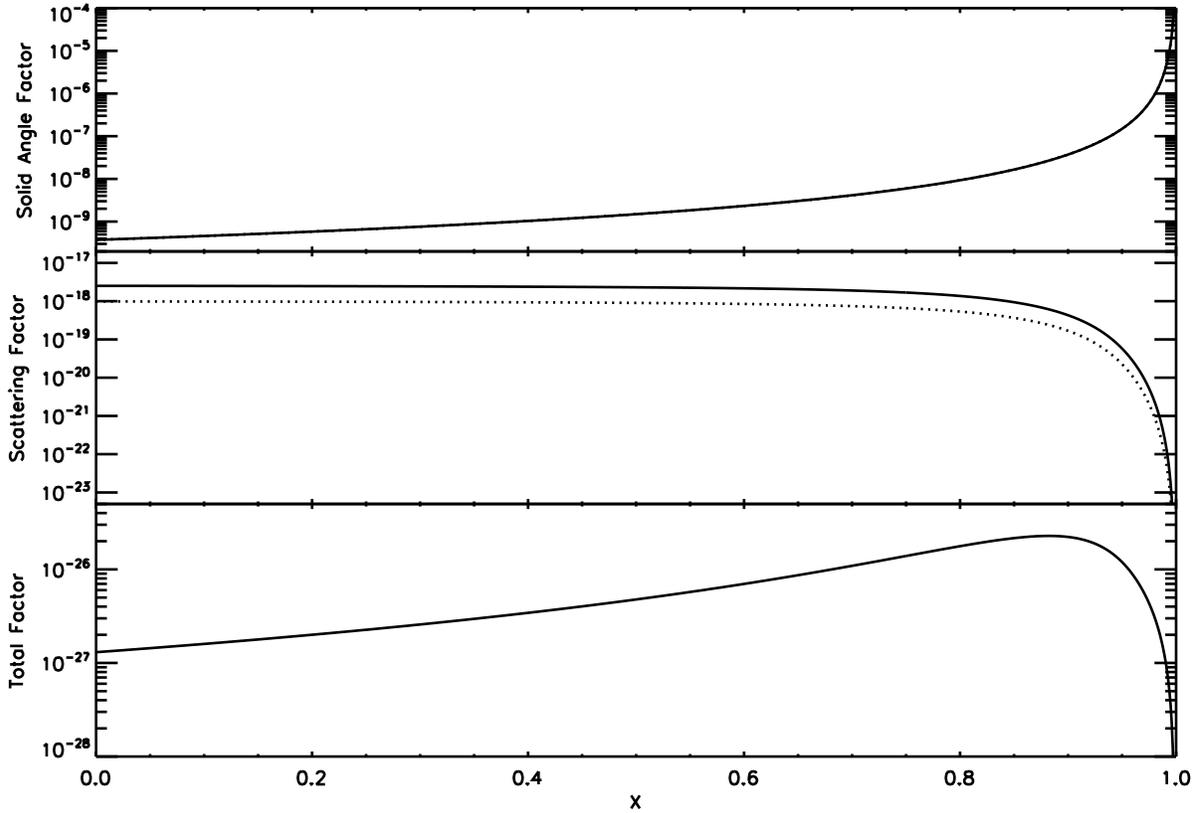}
\caption{Plots of last two terms and their product of the flux ratio (see equation 4) as a function of x.  {\it Top:} 
Plot of the solid angle term which is a measure of the flux the feature intercepts.  {\it Middle:} Plot is of the scattering terms which takes into account what 
fraction of the X-ray flux is scattered to the observer. The solid lines is for silicates scatters and the dotted line is for graphite scatters. {\it Bottom:} This final plot is 
the product of the solid angle term with the sum of the two scattering terms.}
\end{figure}

\begin{figure}
\includegraphics[angle=90,scale=.7]{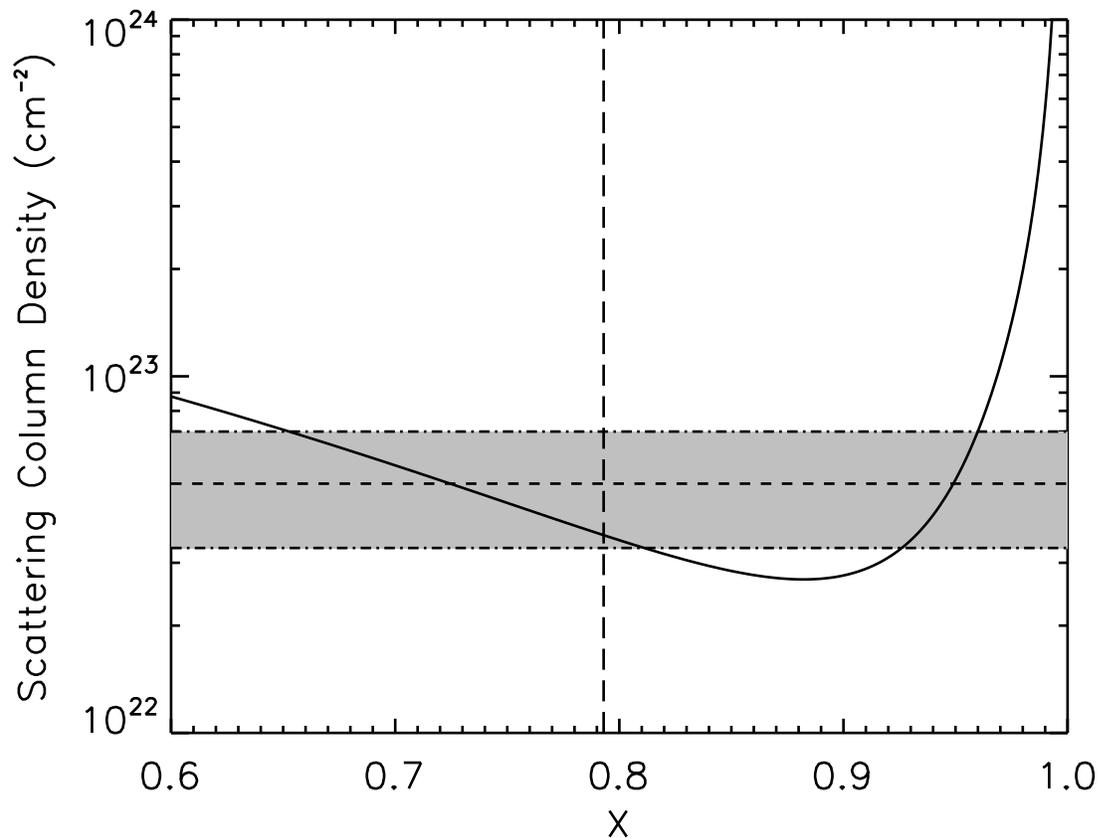}
\caption{A plot of the scattering column density necessary to produce the observed feature/Cygnus X-3 flux ratio (see equation 4) vs x.  
The vertical line (large dashes) is location for the feature determined from the time delay.  The horizontal dashed lines represent the
scattering column density determined from spectral fit and the shaded region between the dashed-dotted lines represent it uncertainties.  We find good agreement 
between values found for the flux ratio and the spectra fit.}
\end{figure}

\clearpage


\begin{deluxetable}{cccccccc}
\tabletypesize{\footnotesize}
\rotate
\tablecolumns{8}
\tablewidth{0pc}
\tablecaption{Cygnus X-3 HETG Chandra Observations \label{tab1}}
\tablehead{ObsID & Date (MJD) & Instrument & Data Mode  & Exp Mode & Exposure (ksec) & Count $\rm Rate^a$ (cts/sec)& State}
\startdata
\dataset [ADS/Sa.CXO#obs/101]{$\rm 101^b$} &  51471 & ACIS-S & FAINT & TE & 1.95 & 35.8 & t/mf \\
\dataset [ADS/Sa.CXO#obs/1456]{1456 $\rm (obi~0)^b} $ & 51471  & ACIS-S & FAINT & TE & 12.12 & 27.7 & t/mf \\
\dataset [ADS/Sa.CXO#obs/1456]{1456 (obi 2)} & 51531 & ACIS-S & FAINT & TE & 8.42 & 18.0 & q/qi \\
\dataset [ADS/Sa.CXO#obs/425]{$\rm  425^c$} & 51638 & ACIS-S & GRADED & TE & 18.54 & 88.2 & fs/mrf \\
\dataset [ADS/Sa.CXO#obs/426]{$\rm  426^c$} & 51640 & ACIS-S &  GRADED & TE & 15.68  & 71.0 & fi/mrf \\
\dataset [ADS/Sa.CXO#obs/6601]{6601} & 53761 & ACIS-S & FAINT & TE & 49.56 & 106.6 & h/qu \\
\enddata
\tablecomments{The states are given as {\it k/s} where: {\it k} are those of \citet{kk} [q: Quiescent, t: Transition, fh: FHXR, fi: FIM, fs:
FSXR, and h: Hypersoft] and {\it s} are those of \citet{as} [qi: Quiescent, mf: Minor flaring, su: Suppressed, qu:
quenched, mrf: Major flaring, and pf: Post flaring]. }
\tablenotetext{a}{This is calculated by taking the total number of 1-8 keV events in the observation and dividing it by the exposure.}
\tablenotetext{b}{The observation end time of obsid 101 is identical to the observation start time of 1456 (obi 0).  For the purpose of
the analysis in this paper these two obsids will taken to be a single observation.}
\tablenotetext{c}{Obsids 425 and 426 were done in alternating exposure mode.  The shorter frame time observation (0.3 sec) resulted in an
additional 0.4 ksec (425) and 0.3 ksec (426) exposure.  For this analysis only the observations with the longer frame time (1.8 sec) were used.}
\end{deluxetable}



\begin{deluxetable}{ccccc}
\tablecolumns{5}
\tablewidth{0pc}
\tablecaption{Phase Folded Light Curve Fitting Parameters \label{tab2}}
\tablehead{Model & Binning & Amplitude & Phase Shift  & $\rm \chi^2 / dof $ }
\startdata
 & 10 & $\rm 0.29 \pm 0.01$ & $\rm 0.60 \pm 0.10$ & (9.24/8) \\
 & 20 & $\rm 0.29 \pm 0.01$ & $\rm 0.55 \pm 0.05$ & (23.74/18) \\   
Phase Shift & 40 & $\rm 0.29 \pm 0.01$ & $\rm 0.55 \pm 0.03$ & (50.82/38) \\  
 & 50 & $\rm 0.29 \pm 0.01$ & $\rm 0.56 \pm 0.02$ & (51.17/48)  \\
 & 100 &  $\rm 0.29 \pm 0.01$ & $\rm 0.56 \pm 0.02$ & (104.60/98)  \\
\tableline
 & 10 & $\rm 0.069 \pm 0.002$ & \nodata & (66.34/9) \\
 & 20 & $\rm 0.068 \pm 0.002$ & \nodata & (80.81/19) \\   
Constant & 40 & $\rm 0.068 \pm 0.002$ & \nodata & (108.07/39) \\  
 & 50 & $\rm 0.068 \pm 0.002$ & \nodata & (110.38/49)  \\
 & 100 & $\rm 0.067 \pm 0.002$  & \nodata & (166.39/99)  \\

\enddata
\tablecomments{The Phase shifted model is the one give in Equation 1.  The Constant model is the same as
the Phase shifted model except that the shift term has been replaced with a constant which is used as a
fit parameter.}
%
\end{deluxetable}



\begin{deluxetable}{cccccc}
\tablecolumns{6}
\tablewidth{0pc}
\tablecaption{Spectral Fit Parameters for the Feature \label{tab3}}
\tablehead{Obsid: & 101+1456(0) & 1456(2) & 425  & 426 & 6601 }
\startdata

$\rm Net Counts^{a}:$ & 163 & 78 & 750 & 697 & 3216 \\
\tableline
Absorbed Power Law: & & & & & \\
\tableline
$\rm N_h~(10^{22}~cm^{-2})$ & $\rm 7.3^{+16.7}_{-7.3}$ & $\rm 6.0^{+36.6}_{-5.5}$  & $\rm 12.5^{+6.7}_{-4.2}$  & $\rm 13.3^{+6.9}_{-4.5}$ & $\rm 10.6^{+2.1}_{-1.8}$ \\
$ \Gamma $ &  $\rm 1.7^{+4.2}_{-2.9}$ & $\rm 2.8^{+4.5}_{-2.8}$ & $\rm 4.8^{+2.0}_{-1.5}$  & $\rm 3.8^{+1.6}_{-1.2}$  & $\rm 4.0^{+0.6}_{-0.6}$ \\
$\rm Flux^{b}$ & $\rm 6.6 \times 10^{-13}$ & $\rm 3.6 \times 10^{-13}$ & $\rm 2.6 \times 10^{-12}$  &  $\rm 2.6 \times 10^{-12}$ & $\rm 3.7 \times 10^{-12}$ \\
$\rm \chi^2 / dof $ &  5.8/8 & 7.7/13  & 14.5/13  & 17.2/13 & 15.2/17 \\
\tableline
Absorbed Blackbody: & & & & & \\
\tableline
$\rm N_h~(10^{22}~cm^{-2})$ & $\rm 1.8^{+13.3}_{-1.8}$ & $\rm 2.3^{+24.4}_{-2.2}$  & $\rm 7.3^{+4.5}_{-2.7}$ & $\rm 7.5^{+4.5}_{-2.8}$ &  $\rm 5.6^{+1.3}_{-1.1}$ \\
$\rm T ~(keV)$ & $\rm 2.4^{+0.9}_{-1.6}$ & $\rm 1.1^{+1.6}_{-0.6}$ &  $\rm 0.7^{+0.2}_{-0.2}$ & $\rm 1.0^{+0.3}_{-0.2}$ & $\rm 0.9^{+0.1}_{-0.1}$ \\
$\rm Flux^{b}$ & $\rm 7.1 \times 10^{-13}$ &  $\rm 3.8 \times 10^{-13}$ & $\rm 2.5 \times 10^{-12}$ & $\rm 2.6 \times 10^{-12}$  & $\rm 3.5 \times 10^{-12}$ \\
$\rm \chi^2 / dof $ & 5.9/9 &7.3/13  & 13.6/13  & 16.4/13  & 15.5/17 \\

\enddata
%
\tablenotetext{a}{Net number of counts in 1-8 keV part of the spectrum.}
\tablenotetext{b}{Measured flux in $\rm ergs~sec^{-1}~cm^{-2}$ in the 1-8 keV band.}
\end{deluxetable}



\begin{deluxetable}{cccccc}
\tablecolumns{6}
\tablewidth{0pc}
\tablecaption{Continuum Spectral Fit Parameters for Cygnus X-3 \label{tab4}}
\tablehead{Obsid: & 101+1456(0) & 1456(2) & 425  & 426 & 6601 }
\startdata
$\rm N_h~(10^{22}~cm^{-2})$ & $\rm 4.12^{+0.17}_{-0.16}$ & $\rm 6.18^{+0.44}_{-0.42}$ & $\rm 3.41^{+0.25}_{-0.26}$ & $\rm 4.34^{+0.47}_{-0.44}$ & $\rm 2.75^{+0.11}_{-0.10}$ \\
$ \Gamma $ & $\rm 0.72^{+0.05}_{-0.05}$ & $\rm 0.63^{+0.09}_{-0.09}$ & \nodata &  \nodata & \nodata \\
$\rm T ~(keV)^{a}$ & \nodata & \nodata & $\rm 2.25^{+0.04}_{-0.04}$ & $\rm 2.22^{+0.05}_{-0.04}$ & $\rm 2.00^{+0.02}_{-0.02}$ \\ 
$\rm N_{pc}~(10^{22}~cm^{-2})$  & \nodata  & \nodata  & $\rm 3.91^{+0.28}_{-0.24}$ & $\rm 4.58^{+0.36}_{-0.36}$ & $\rm 3.22^{+0.22}_{-0.19}$ \\
$\rm f_{pc} $  & \nodata  & \nodata  & $\rm 0.74^{+0.06}_{-0.06}$ & $\rm 0.85^{+0.04}_{-0.06}$ & $\rm 0.62^{+0.04}_{-0.04}$ \\
$\rm Flux^{b}$  & $\rm 1.69 \times 10^{-9}$ & $\rm 1.11 \times 10^{-9}$ & $\rm 5.86 \times 10^{-9}$ & $\rm 4.95 \times 10^{-9}$ & $\rm 6.13 \times 10^{-9}$ \\
$\rm \chi^2 / dof $  & 4151.55/8660 & 1668.70/4319 & 3269/4317 & 2502.15/4317 & 6016.95/4317 \\
\enddata
\tablecomments{The {\it Chandra} grating spectral are rich in spectral features \citep{fp}.  To achieve acceptable fits we included a large number of spectral features 
(emission lines, absorption lines, edges, and radiative recombination continuum).}
\tablenotetext{a}{Temperature of a disk blackbody.}
\tablenotetext{b}{Measured flux in $\rm ergs~sec^{-1}~cm^{-2}$ in the 1-8 keV band.}
\end{deluxetable}



\begin{deluxetable}{cccccc}
\tablecolumns{6}
\tablewidth{0pc}
\tablecaption{Spectral Fit Parameters for Scattering Model for the Feature \label{tab5}}
\tablehead{Obsid: & 101+1456(0) & 1456(2) & 425  & 426 & 6601 }
\startdata
$\rm N_{cl}~(10^{22}~cm^{-2})$ & $\rm 6.4^{+15.5}_{-4.19}$ & $\rm 0.0^{+7.8}_{-0.0}$ & $\rm 3.1^{+1.6}_{-1.3}$ & $\rm 4.0^{+6.4}_{-4.1}$ & $\rm 5.0^{+2.0}_{-1.7}$ \\
$\rm A $ & $\rm 1.4^{+0.5}_{-0.5}$ & $\rm 1.5^{+0.7}_{-0.7}$ & $\rm 1.1^{+0.2}_{-0.2}$ &  $\rm 1.1^{+0.2}_{-0.2}$ &  $\rm 1.4^{+0.1}_{-0.1}$ \\
$\alpha$ & $\rm 2.0^{*}$  & $\rm 2.0^{*}$  & $\rm 2.0^{*}$ & $\rm 1.8^{+1.6}_{-1.2}$ & $\rm 1.9^{+0.6}_{-0.6}$ \\ 
$\rm Flux^{a}$  & $\rm 6.0 \times 10^{-13}$ & $\rm 3.9 \times 10^{-13}$ & $\rm 2.7 \times 10^{-12}$ & $\rm 2.6 \times 10^{-12}$ & $\rm 3.6 \times 10^{-12}$ \\
$\rm \chi^2 / dof $  & 6.4/8 & 7.8/13 & 15.8/13 & 16.7/12 & 14.7/16 \\
\enddata
%
\tablenotetext{*}{Fixed.}
\tablenotetext{a}{Measured flux in $\rm ergs~sec^{-1}~cm^{-2}$ in the 1-8 keV band.}
\end{deluxetable}



\begin{deluxetable}{cc}
\tablecolumns{2}
\tablewidth{0pc}
\tablecaption{Scattering Flux Calculation Parameters \label{tab6}}
\tablehead{Parameter & Value }
\startdata

$\rm F_{cx3}$~(1-8~keV)& $\rm 9.12 \times 10^{-1} ~photons ~cm^{-2} s^{-1}~~(6.13 \times 10^{-9} ~ergs ~cm^{-2} s^{-1})$ \\
$\rm F_{lf}$~(1-8~keV) & $\rm 5.6 \times 10^{-4} ~photons ~cm^{-2} s^{-1}~~(3.6 \times 10^{-12} ~ergs ~cm^{-2} s^{-1})$ \\
$\theta$ & $\rm 15.6^{\prime\prime}$ \\
$\alpha_1$ & $\rm 2.77^{\prime\prime}$  ~(5 Sigma Full Width:~ $5.54^{\prime\prime}$) \\
$\alpha_2$ & $\rm 1.82^{\prime\prime}$  ~(5 Sigma Full Width:~ $3.65^{\prime\prime}$) \\
$\rm D$ & 9 kpc \\
$\rho ~(graphite)^a $ &   $\rm 2.24 ~g~cm^{-3}$ \\
$\rho ~(silicate)^a $ &   $\rm 3.5 ~g~cm^{-3}$ \\
$\rm N_d ~(graphite)^a $ &  $\rm 7.41 \times 10^{-16} ~\frac{grains}{H~atom~\mu}$ \\
$\rm N_d  ~(silicate)^a $ & $\rm 7.76 \times 10^{-16} ~\frac{grains}{H~atom~\mu}$ \\
$\rm a_{min}~^a$ & $0.005 ~\mu$ \\
$\rm a_{max}~^a$ & $0.25 ~\mu$ \\
$\rm E_1$ & 1.0 keV \\
$\rm E_2$ & 8.0 keV \\
\enddata
%
\tablenotetext{a}{The dust grain parameters were taken from \citet{wd}.}
\end{deluxetable}


\end{document}